\documentclass{article}
\usepackage{spconf,amsmath,amssymb,graphicx,booktabs}
\usepackage[nocompress,noadjust]{cite}
\usepackage[hidelinks]{hyperref}
\usepackage{xurl}
\usepackage{eso-pic}
\hypersetup{pdftitle={Studying Detection Rule Generation as a Unified Task},pdfauthor={Cheng Meng et al.},pdfkeywords={Detection rule generation, large language models, retrieval augmented generation, automated evaluation}}

\title{STUDYING DETECTION RULE GENERATION AS A UNIFIED TASK}
\name{\shortstack{Cheng Meng$^{1,2}$, Wenxin Le$^{1,2}$, Xinyi Li$^{1,2}$, Qiuyun Wang$^{1,2}$,\\
Fangli Ren$^{1,*}$, Zhengwei Jiang$^{1,2}$, Baoxu Liu$^{1,2}$}\thanks{$^{*}$Corresponding author. Email: \href{mailto:renfangli@iie.ac.cn}{renfangli@iie.ac.cn}}}
\address{$^{1}$Institute of Information Engineering, Chinese Academy of Sciences, Beijing, China\\
$^{2}$School of Cyber Security, University of Chinese Academy of Sciences, Beijing, China}

\begin{document}

\AddToShipoutPictureFG*{%
  \AtPageLowerLeft{%
    \put(\LenToUnit{0.75in},\LenToUnit{0.25in}){%
      \parbox[b]{7in}{\normalfont\fontsize{9}{10.5}\selectfont
        This work has been submitted to the IEEE for possible publication. Copyright may be transferred without notice, after which this version may no longer be accessible.}%
    }%
  }%
}
\ninept
\renewcommand{\footnotesize}{\fontsize{9}{10.5}\selectfont}
\maketitle

\begin{abstract}
Security systems use detection rules to identify suspicious activity. Existing studies often investigate rule generation for specific security systems, devoting substantial effort to developing dedicated methods and evaluation setups. Such customization contributes to fragmented research, limiting method reuse and result comparability across systems. We therefore study detection rule generation as a unified task across diverse natural language inputs and rule languages. To support method reuse, we propose UniRule, which abstracts diverse rules into shared natural language representations for retrieval. To enable consistent evaluation, we introduce a protocol that compares rules under shared criteria and aggregates the results into method scores. Experiments demonstrate the effectiveness of UniRule and the reliability of the evaluation protocol. They also show that method performance in one setting can be predicted from results in others, with average error close to that obtained using that setting's own data. These findings support studying detection rule generation as a unified task.

\end{abstract}


\section{Introduction}
\label{sec:intro}

Detection rules express detection requirements as conditions that security systems can evaluate. Writing them manually requires substantial effort and security expertise, motivating automated generation~\cite{wang2026rulepilot}. Recent studies use large language models (LLMs) to generate detection rules from natural language descriptions and threat reports~\cite{wang2026rulepilot,schwartz2025llmcloudhunter,mitra2025falcon,cai2026texts,ghaffarzadegan2026threat,chakraborty2026cti}. Although inputs and rule languages differ across applications, these studies often adopt similar approaches to rule generation, such as rule retrieval and refinement through feedback~\cite{wang2026rulepilot,mitra2025falcon,ghaffarzadegan2026threat,li2025gridai}.

Existing studies often implement these similar generation approaches using workflows and resources tailored to specific security scenarios. Information extraction focuses on cloud API calls~\cite{schwartz2025llmcloudhunter}, shell commands~\cite{cai2026texts}, or features of network payloads~\cite{li2025gridai}, depending on the application. Retrieval uses indexes of Sigma templates~\cite{ghaffarzadegan2026threat} or models trained to match threat descriptions with Snort and YARA rules~\cite{mitra2025falcon}. Feedback for rule refinement comes from Sigma rule checks~\cite{ghaffarzadegan2026threat}, execution in Splunk~\cite{wang2026rulepilot}, or tests on network traffic~\cite{li2025gridai}. These choices serve the needs of their target applications. Reusing the same approaches in a different scenario often requires preparing suitable resources and adapting the implementation.

This application focus also shapes the evaluation of generated rules. Execution tests use logs~\cite{wang2026rulepilot}, network traffic~\cite{li2025gridai}, or program samples~\cite{li2023packgenome} to evaluate how well rules detect the target behaviors. The test samples and expected outcomes must reflect the detection requirements~\cite{gao2021enabling}. Preparing and verifying them requires substantial data collection and expert effort~\cite{bertiger2025evaluating,uetz2024you,virkud2026credible}. Semantic evaluation also involves preparing data and scoring tools for the target application. Reference rules or expert annotations specify the detection content expected for each application, such as cloud API calls~\cite{schwartz2025llmcloudhunter,ghaffarzadegan2026threat} or command patterns~\cite{cai2026texts}. Preparing the annotations requires expert interpretation of threat reports~\cite{schwartz2025llmcloudhunter}. Another approach trains a scorer on pairs of rules in the target languages and generated threat descriptions~\cite{mitra2025falcon}. These evaluations provide evidence for their target applications, but differences in the data and scoring criteria make the reported scores difficult to compare directly.

These studies have advanced detection rule generation and its evaluation, yet research remains fragmented across security applications. Even when approaches are similar, substantial effort is needed to develop methods and establish their effectiveness in each application. These demands are particularly difficult to meet when suitable test data and evaluation resources are scarce. To address this fragmentation, we study detection rule generation as a unified task, aiming to support method reuse and consistent evaluation across natural language input forms and rule languages.

To support method reuse, we introduce UniRule, showing that the same method can improve rule generation across these settings. It uses an LLM to generate natural language descriptions of what each existing rule detects and how it detects it. By matching inputs to these descriptions, UniRule retrieves rules in the same way across natural language input forms and rule languages. The retrieved rules then guide generation.

To enable consistent evaluation, we propose a general protocol for comparing generation methods. Meng et al.~\cite{meng2026can} provide evidence of LLMs' ability to compare detection rules in pairs. Our protocol uses this ability to compare rules generated for the same natural language input. A Bradley--Terry model aggregates these comparisons into method scores. The same criteria and scoring method apply across input forms and rule languages. The protocol requires no reference rules, execution data, or additional training of the LLM judge.

Experiments across three rule languages and three input forms show consistent benefits from rule examples and retrieval, with UniRule ranking highest among the four generation methods in all nine settings. To assess the protocol's reliability, we compared its judgments with those of three security experts and found agreement comparable to that among the experts. We also predict method performance using results from settings that share neither the target rule language nor its input form. The average prediction error is close to that obtained using the target setting's own data.

Toward unified detection rule generation, we make the following contributions.\footnote{Code is available at \url{https://github.com/MengC1024/UniRule}.} (1)~We introduce UniRule, which improves rule generation across settings using the same semantic retrieval method. (2)~We develop a general evaluation protocol that compares rules under shared criteria and aggregates judgments into method scores, with reliability assessed through expert agreement. (3)~We provide empirical support for studying detection rule generation as a unified task. Rule examples and retrieval yield consistent benefits across settings, and results from other settings predict method performance.

\section{Unified Rule Generation}
\label{sec:method}

\subsection{Task definition}

Given a detection context expressed in natural language and a target rule language, the task is to generate a rule that satisfies the requirements expressed in the context. We consider contexts that vary in content and form, and target languages with different rule formats and detection mechanisms. Each combination of an input form and a rule language defines a setting.

\subsection{Rule retrieval in UniRule}

\begin{figure}[!htbp]
\centering
\includegraphics[trim={0 6bp 0 2bp},clip]{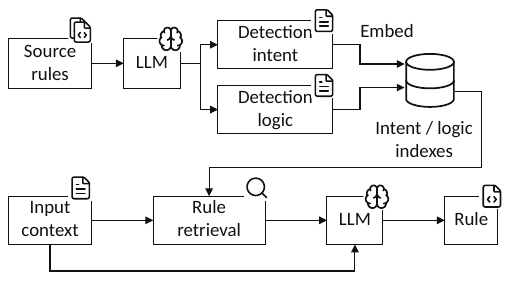}
\caption{Overview of UniRule. An LLM generates detection intent and logic descriptions of source rules. These descriptions are embedded and indexed to retrieve rules relevant to the input context. The retrieved rules and input context guide an LLM to generate a detection rule.}
\label{fig:unirule}
\end{figure}

To retrieve examples for a detection requirement, UniRule describes existing rules through two shared semantic aspects, detection intent and detection logic (Fig.~\ref{fig:unirule}). Detection intent identifies the behavior a rule is meant to detect. Detection logic describes the observable conditions and how they are combined to detect that behavior. Expressing both in natural language allows UniRule to retrieve rules by matching their descriptions to the input. The source rules provide concrete implementations for generation.

Offline, an LLM generates intent and logic descriptions for each source rule. These descriptions are then embedded and indexed separately, with each entry linked to its source rule.

For illustration, consider a Splunk rule that detects the execution of files with double extensions. Its \emph{intent} is ``Detect executables that use double extensions to masquerade as document files.'' Its \emph{logic} is ``Search process events for \texttt{Processes.process} values matching extension patterns such as \texttt{*.doc.exe} or \texttt{*.pdf.exe}.''

UniRule encodes the input context with the same embedding model. It compares this embedding with the intent and logic embeddings of source rules in the target language using cosine similarity. Inputs differ in how much they specify intent and logic, so UniRule ranks each rule by the higher of the two similarities. This prevents a weaker match in one aspect from lowering the retrieval score. An LLM then generates a rule using the input context, target language, and the top $k$ distinct rules.

This design allows UniRule to handle different input forms and rule languages without changing its retrieval method.

\section{Evaluating Rule Generation Methods}
\label{sec:evaluation}

\subsection{Pairwise rule comparison}

An LLM judge compares two candidate rules for the same input context and target language. Method names are hidden from the judge. It first identifies the detection requirements in the input and assesses each rule against them. It then compares the candidates by faithfulness to the context, technical correctness, and absence of unsupported additions, in that order.

Faithfulness concerns whether the required behavior and conditions are retained. Technical correctness concerns the rule's constructs and use of fields and operators. The final criterion penalizes additional conditions without support in the context, since they can change what the rule detects. Across settings, we use the same criteria and priority order to assess how well rules meet the input requirements. The judge is instructed not to favor rules merely for their length or metadata; ties are allowed when neither candidate has a meaningful advantage.

The judge uses the input context, target language, and candidate pair, without reference answers or execution data.

\subsection{Method scores}

For each input, we compare every pair of candidate rules. We randomize presentation order and balance A and B positions across inputs~\cite{chiang2024chatbot}. For scoring, we link each outcome to the methods that supplied the two rules. Judgments from different inputs therefore provide repeated comparisons between the same methods. A Bradley--Terry (BT) model~\cite{bradley1952rank} combines this evidence into one score per method.

Position bias has been documented in LLM judging~\cite{zheng2023judging,liusie2024llm}. In addition to balancing presentation order, we use an extension of BT that models the position effect separately from method scores~\cite{turner2012bradley}.

Let $\xi_i$ denote the score for method $i$. In a comparison, methods $i$ and $j$ supply the rules presented as A and B. We record an A win as $Y=1$, a tie as $Y=0.5$, and a B win as $Y=0$. A tie thus counts as half a win for each method. The model predicts the expected outcome from the score difference $\xi_i-\xi_j$, with an additional term $\gamma$ for the effect of presentation as A,
\begin{equation}
 \widehat{Y}_{ij}=\frac{1}{1+\exp[-(\xi_i-\xi_j+\gamma)]}.
 \label{eq:bt}
\end{equation}
The prediction $\widehat{Y}_{ij}$ represents the win rate of method $i$ against method $j$, with ties counted as above. A positive $\gamma$ indicates a tendency to favor A.

We jointly fit the method scores and $\gamma$ using logistic regression to explain the observed wins, losses, and ties across method pairs. Each method's score reflects its results against the other methods across the inputs included in that fit. Because method scores enter Eq.~\ref{eq:bt} only through their differences, we impose an identifiability constraint by fixing one method's score at zero.

We calculate reported method win rates from the fitted score differences, excluding the estimated effect of presentation order. To do this, we retain the fitted method scores and set $\gamma=0$ in Eq.~\ref{eq:bt}. Two methods with equal scores then have a predicted win rate of $0.5$.

For each setting, we fit a separate model using only comparisons from that setting. To obtain overall method scores, we pool comparisons from all settings and fit one model to the combined data.

\section{Experiments}
\label{sec:experiments}

\subsection{Experimental setup}

Detection requirements can describe the behavior to detect~\cite{wang2026rulepilot} or specify the conditions for detecting it, a distinction that also underlies UniRule. Requirements covering diverse input forms and rule languages are difficult to obtain. We therefore build the inputs from rules written for practical detection needs. We synthesize Intent and Logic inputs to expand coverage, retaining the rules' original descriptions as Context. These natural language inputs can be checked by human reviewers against the source rules. We sample 100 test rules each from Splunk Security Content (SPL),\footnote{\url{https://github.com/splunk/security_content}} Elastic queries,\footnote{\url{https://github.com/elastic/detection-rules}} and Snort signatures,\footnote{\url{https://www.snort.org/downloads}} excluding them from retrieval. These yield 900 inputs across nine settings, with 100 inputs per setting. We compare the same generation methods under shared evaluation criteria across these settings.

We compare Direct with three retrieval augmented generation (RAG) methods. Direct generates rules from the input context and target language without examples. Each RAG method provides 15 example rules in the target language. Random RAG selects rules randomly. Standard RAG retrieves by similarity between the input and raw rule text. UniRule retrieves through intent and logic descriptions. For each input, all methods share the generator and task instructions.

For dataset construction, DeepSeek-V3.2 in non-thinking mode generates Intent and Logic inputs from the test rules. All these inputs are then manually reviewed and revised as needed. Qwen3.8-27B-FP8 generates detection rules for all four methods. It also generates intent and logic descriptions of source rules (for retrieval) in UniRule. Qwen3-Embedding-8B provides embeddings for retrieval~\cite{zhang2025qwen3}. For evaluation, each input has five candidate rules, comprising four generated rules and the original human rule. GPT-5.6 Sol compares all ten candidate pairs using the protocol in Section~\ref{sec:evaluation}, yielding 9,000 judgments with balanced presentation order.

BT significance tests and 95\% Wald confidence intervals use standard errors clustered by original rule~\cite{colin2015practitioner}. We apply Holm correction to the prespecified tests within each fit. For expert agreement and prediction errors, we obtain 95\% bootstrap confidence intervals by resampling original rules within each dataset, keeping their input forms and comparisons together.

\subsection{Agreement with experts}

To assess the protocol's reliability, three security experts independently judged the same 270 rule pairs from 90 original rules across all settings and method pairs. The experts used the same evaluation criteria as the LLM. Method identities and LLM judgments were hidden from the experts. We report exact agreement and Cohen's $\kappa$ to account for chance agreement.

\begin{figure}[!htb]
\centering
\includegraphics[trim={0 3bp 0 2bp},clip]{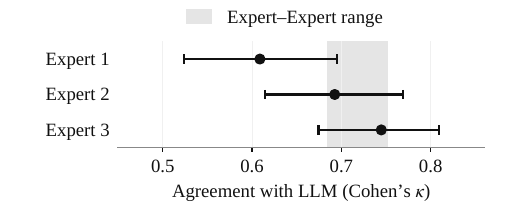}
\caption{Agreement with experts. Points show agreement with the LLM, with 95\% bootstrap confidence intervals. Shading spans the three $\kappa$ estimates between experts.}
\label{fig:agreement}
\end{figure}

Agreement with the LLM is 76.7--85.9\% ($\kappa=0.609$--$0.745$), compared with 81.1--85.2\% ($\kappa=0.685$--$0.751$) between experts. Agreement with the LLM falls within this $\kappa$ range for two experts and below it for Expert 1 (Fig.~\ref{fig:agreement}). These results support the reliability of the protocol for comparing generation methods.

\subsection{Method effectiveness across settings}

The four generation methods rank in the same order in all nine settings, with UniRule highest (Table~\ref{tab:results}). Its adjusted win rate against Standard RAG is 59.6\% overall (95\% CI 56.9--62.2\%), with significant advantages for each rule language after pooling input forms and applying Holm correction ($p<0.01$). It improves generation across input forms and rule languages with the same core design.

\begin{table}[!htb]
\centering
\caption{BT scores by setting and overall (Direct $=0$). R/S/U/H denote Random RAG, Standard RAG, UniRule, and human reference rules.}
\label{tab:results}
\setlength{\tabcolsep}{4pt}
\begin{tabular}{@{}llrrrr@{}}
\toprule
Rule language & Input & R & S & U & H \\
\midrule
Elastic & Context & 1.21 & 1.27 & 1.79 & 2.61 \\
        & Intent  & 1.13 & 1.82 & 1.98 & 3.70 \\
        & Logic   & 1.47 & 2.19 & 2.36 & 3.77 \\
Snort   & Context & 0.87 & 2.10 & 2.28 & 2.72 \\
        & Intent  & 1.06 & 1.76 & 2.18 & 2.75 \\
        & Logic   & 2.48 & 2.63 & 3.44 & 3.93 \\
Splunk  & Context & 0.57 & 0.99 & 1.43 & 2.38 \\
        & Intent  & 0.92 & 1.50 & 1.85 & 3.48 \\
        & Logic   & 1.46 & 1.96 & 2.46 & 3.71 \\
\midrule
\multicolumn{2}{l}{Overall} & 1.17 & 1.72 & 2.11 & 3.08 \\
\bottomrule
\end{tabular}
\end{table}

\begin{figure}[!t]
\centering
\includegraphics{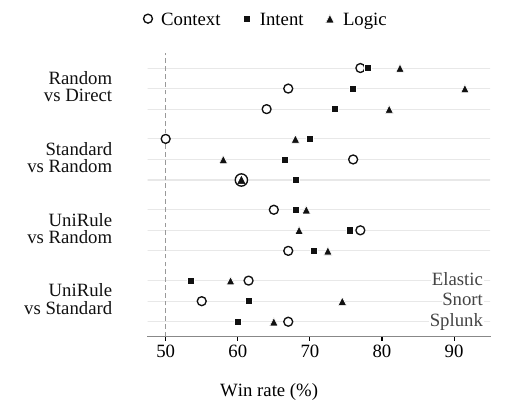}
\caption{Benefits of rule examples and retrieval across settings. Points show observed win rates of the first named method, counting ties as half a win. Within each comparison, rows show Elastic, Snort, and Splunk from top to bottom. Overlapping markers indicate equal values.}
\label{fig:outcomes}
\end{figure}

Figure~\ref{fig:outcomes} shows that providing rule examples improves generation, and retrieval generally brings further gains. Random RAG versus Direct assesses the benefit of providing examples, while the other comparisons examine how example selection affects performance. These benefits vary in magnitude across rule languages and input forms, but show a consistent overall pattern.

\subsection{Predicting performance in new settings}

We test whether results from other settings predict how often one method outperforms another~\cite{wainer2023bayesian}. Each setting serves as the target. The 8:1 scheme uses all eight other settings as sources. The 4:1 scheme uses only the four settings that share neither the target rule language nor its input form. For example, when Snort with Intent is the target, the four source settings under 4:1 are Context and Logic from Elastic and Splunk.

We focus on the three RAG methods to test whether their smaller performance differences can be predicted from other settings. We fit the BT model in Section~3.2 using source comparisons. Keeping the fitted scores and position term fixed, we apply Eq.~\ref{eq:bt} to target method pairs and their presentation order. Target outcomes are used only to evaluate these source predictions. We repeat tenfold cross validation ten times, holding out one fold within each rule language in every split and keeping all input forms of a rule together.

We compare BT with three baselines. The equal score baseline assumes no performance differences between methods. The ranking baseline fits a shared win rate for higher versus lower ranked methods, ignoring the size of score differences. The pairwise baseline estimates a separate win rate for each method pair. All are fitted on source comparisons and account for presentation order. As a reference, we fit BT using comparisons from the target setting's training inputs to predict performance on its test inputs. Each fold uses 720 source inputs under 8:1, 360 under 4:1, or 90 target inputs, and tests all predictors on the same ten target inputs. We report mean squared error (MSE) between predicted and observed comparison outcomes, averaged over target settings and repetitions. Confidence intervals are obtained by rerunning the full analysis on bootstrap samples.

\begin{figure}[!t]
\centering
\includegraphics[trim={0 4bp 0 6bp},clip]{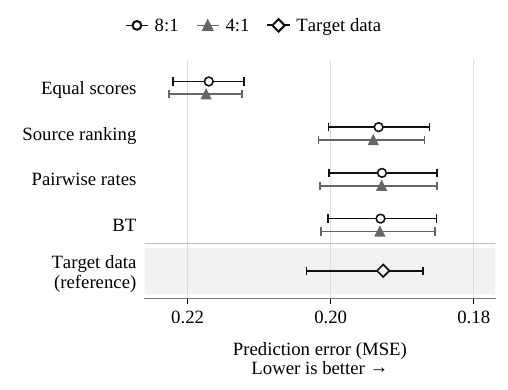}
\caption{Mean prediction error over nine target settings and ten repetitions. Lower MSE appears farther right. Intervals are 95\% bootstrap confidence intervals. Source predictors use 8:1 or 4:1; the target data reference uses training inputs from the target setting.}
\label{fig:transfer}
\end{figure}

All three source predictors improve on equal scores and have similar mean MSE (Fig.~\ref{fig:transfer}). Compared with equal scores, BT reduces mean MSE by 11.1\% under 8:1 and 11.2\% under 4:1. BT has a mean MSE of 0.1930 under 8:1 and 0.1931 under 4:1, both close to 0.1927 for the target data reference. Both paired differences from the reference have 95\% confidence intervals including zero. These results show that other settings provide useful predictions even when sources exclude the target rule language and input form. Training data sizes differ, so these results do not establish equivalent performance.

\section{Discussion \& Limitations}
\label{sec:discussion}

Our inputs are based on existing rules developed for practical detection needs. Although all synthetic inputs have been reviewed by humans, the abstracted requirements may still differ from those specified in practice. Moreover, our experiments assess how well generated rules satisfy the supplied requirements but do not measure detection effectiveness in deployment.

The synthetic inputs share the semantic abstraction used by UniRule, which may influence comparisons between generation methods. Nevertheless, all four methods follow the same ranking on the original rule descriptions (nonsynthetic inputs) and synthetic inputs in each of the three rule languages (Table~\ref{tab:results}). UniRule achieves similar gains over Standard RAG on synthetic and nonsynthetic inputs. Pooling comparisons across the three rule languages, its adjusted win rate is 59.4\% for nonsynthetic inputs (95\% CI 55.0\%--63.6\%), compared with 57.6\% and 61.9\% for the two forms of synthetic input. We find no clear indication that the shared abstraction gives UniRule an additional advantage on synthetic inputs.

Our experiments use a single rule generation model. Whether the observed consistency extends to other generation models remains to be tested.

\section{Conclusion}

We studied detection rule generation as a unified task using UniRule and a general evaluation protocol. Across the natural language input forms and rule languages examined, we observe consistency in retrieval benefits and method performance.

In security operations, teams responding to new threats may have limited data for evaluating alternative rule generation methods~\cite{bertiger2025evaluating,uetz2024you}. This consistency provides an empirical basis for using results from other security scenarios to guide method selection under these constraints.

\bibliographystyle{IEEEbib}
\bibliography{custom}

\section{Acknowledgments}

This work was supported by the Youth Innovation Promotion Association, CAS (No.~2023170) and Beijing Key Laboratory of Network Security and Protection Technology. We declare no conflicts of interest.

We used OpenAI Codex for English language polishing and \LaTeX{} typesetting. We reviewed all content and take full responsibility for the manuscript and its results.

\section{Compliance with Ethical Standards}

Three colleagues voluntarily provided expert annotations after being informed of the study's purpose. No personal information was collected during annotation.

\end{document}